\documentclass[aps,twocolumn,pra,showpacs,floatfix]{revtex4}
\usepackage{epsfig}
\usepackage{dcolumn}

\begin{document}

\title{\bf Ionization potentials and polarizabilities of superheavy elements from Db to Cn ($Z$=105 to 112).}

\author{V. A. Dzuba}
\affiliation{School of Physics, University of New South Wales,
Sydney 2052, Australia}

\date{\today}
\begin{abstract}
Relativistic Hartree-Fock and random phase approximation methods for open shells are used 
to calculate ionization potentials and static scalar polarizabilities of eight superheavy elements 
with open $6d$-shell, which include Db, Sg, Bh, Hs, Mt, Ds, Rg and Cn ($Z$=105 to 112).
Inter-electron correlations are taken into account with the use of the semi-empirical polarization potential.
Its parameters are chosen to fit the known ionization potentials of lighter atoms. Calculations 
for lighter atoms are also used to illustrate the accuracy of the approach.

 \end{abstract}
\maketitle
%***************************************************************************

\section{Introduction}

Study of superheavy elements (SHE, nuclear charge $Z>100$)  is a huge and important area of research including 
synthesis of SHE and studying their nuclear structure and atomic and chemical properties (see, e.g. reviews 
\cite{türler2013,schadel2015,hofmann2010,pershina2015,schwerdtfeger2015,eliav2015}). The main focus is on
production and nuclear properties of SHE. All elements with nuclear charge up to $Z=118$ have been
synthesised already~\cite{hofmann2010,oganessian2010}. However, very little experimental data is available 
on atomic properties such as ionization potential, excitation energies, etc.
The most recent achievement of this kind is measurement of ionization 
potential of lawrencium ($Z=103$)~\cite{sato2015measurement} and 
nobelium ($Z=102$)~\cite{Block2015,Sato2015}. 
The energy of the  $7s^2 \ ^1$S$_0 - 7s7p \ ^1$P$_1$ and the hyperfine structure of the $^1$P$_1$ state 
of $^{253}$No has been also measured~\cite{Block2015}.
Information on the atomic properties of heavier elements mostly comes from atomic 
calculations~\cite{schwerdtfeger2015,eliav2015}. Very accurate calculations
are available for SHE with relatively simple electron structure, having one or two of either electrons or 
vacancies on the outermost shells~\cite{eliav2015} or just few valence electrons above closed 
shells~\cite{Dzu13,DSS14,Gd15}. There is a big group of SHE which fell to neither of these categories. These are the 
SHE from dubnium (Db, $Z=105$) to roentgenium (Rg, $Z=111$). They all have the $6d^n7s^2$ ground state 
configuration with $n=3$ for Db and $n=9$ for Rg. The elements have been synthesised and studied experimentally
(see, e.g. \cite{dullmann2002chemical}) but little theoretical data is available on atomic properties of the elements.
This is because complicated electron structure with open $6d$ shell makes accurate calculations difficult 
and approximate approaches should be developed. 

In this paper we present an approach, which allows to make reliable predictions for the values of 
ionization potential and static scalar polarizabilities for atoms with open $d$ or $f$ shell and two $s$ 
valence electrons. 
We demonstrate that for such systems the existence of vacancies in $d$ or $f$ shells are not important 
and the atoms can be treated as closed-shell systems while vacancies are taken into account in an 
approximate way via fractional occupation numbers.
We start the calculations with the relativistic Hartree-Fock method. Then external
electric field is included in the framework of the random-phase approximation (RPA). 
This allows to calculate ionization potential and polarizabilities of noble gases with pretty good accuracy. 
To extend applicability of the method to more
complecated systems and to improve accuracy of the calculations correlations are included semi-empirically via
polarization potential. The potential is scaled to fit the experimental ionization potential, the scaling also leads to
better results for polarizabilities. 

In next section we demonstrate how the method works for closed-shell systems. Then, in last section, we apply it to 
atoms with open $d$ or $f$ shell, including superheavy elements from Db to Cn.

\section{Closed-shell atoms}

We start from a closed-shell atom.
If Hartree-Fock equations for atomic orbitals $\psi_a$ are written in the form
\begin{equation}
	\left( \hat H^{\rm HF} - \epsilon_c\right) \psi_c = 0,
\label{e:HFPsi}
\end{equation}
then equations for the atom in external field in the random-phase approximation (RPA) can be written as
\begin{equation}
	\left( \hat H^{\rm HF} - \epsilon_c\right) \delta\psi_c = -(\hat d +\delta \hat V)\psi_c.	
\label{e:HFDPsi}
\end{equation}
In both equations $ \hat H^{\rm HF} $ is the relativistic Hartree-Fock (HF) Hamiltonian, index $c$ numerates 
single-electron orbitals of the atomic core, $\psi_c$ and $\epsilon_c$ are single-electron HF wave function 
and corresponding energy, $\hat d$ is an operator of external field (in our case it is electric dipole operator, 
$\hat d = -e\hat r$), $\delta\psi_c$ is a correction to atomic orbital $\psi_c$ caused by external field, and 
$\delta \hat V$ is the correction to the self-consistent HF potential caused by external field via changing 
all atomic orbitals. Equations (\ref{e:HFPsi}) and (\ref{e:HFDPsi}) are solved iteratively for all core atomic 
states $c$.

Once iterations are converged, static scalar polarisability can be calculated as
\begin{equation}
\alpha_0 = \frac{2}{3} \sum_{cn} \frac{\langle c||\hat d ||n\rangle\langle c || \hat d + \delta \hat V || n \rangle}
{\epsilon_n - \epsilon_c}
\label{e:alphac}
\end{equation}
Summation goes over core states $c$ and complete set of single-electron states above the core $n$.
Note that correction to the potential $\delta \hat V$ goes to only one of the two electric dipole matrix elements 
in (\ref{e:alphac})~\cite{mitroy2010}.

Expression (\ref{e:alphac}) can be rewritten in a more compact form
\begin{equation}
	\alpha_0 = \frac{2}{3} \sum_c \langle \delta\psi_c||\hat d||\psi_c \rangle.
\label{eq:alpha0}
\end{equation}
This equation is more convenient for calculations since it uses the corrections to the atomic orbitals which
come form solving the RPA equations  (\ref{e:HFDPsi}) and does not require a complete set of single-electron states.

Equation  (\ref{e:HFDPsi})  does not include correlations. However, it gives very accurate results for light
noble gases where correlations are small. Accuracy goes down to few per cent for heavy noble gases 
due to larger correlations. For heavy closed-shell atoms other than noble gases, such as Ba, Hg, Yb, etc., 
the accuracy given by  (\ref{e:HFDPsi}) is very poor due to large correlations. To improve the accuracy we
use a semi-empirical  treatment of correlations.  

It is known that on large distances correlations produce polarization potential of the form
\begin{equation}
  V_p(r) = - \frac{\alpha_0}{2r^4},
\label{e:pol}
\end{equation}
where $\alpha_0$ is atomic polarizability. Potential (\ref{e:pol}) cannot be used on short distances.
It is often replaced in the literature (see, e.g. \cite{Bates1943}) by
\begin{equation}
  V_p(r) = - \frac{a}{2(b^2+r^2)^2},
\label{e:Vp}
\end{equation}
where $b$ is a cut-off parameter introduced to remove singularity at $r=0$ (we use $b=5 a_0$, where $a_0$ 
is Bohr radius), and $a$ is a fitting parameter. Potential $V_p$ (\ref{e:Vp}) is added to the HF Hamiltonian
and the value of $a$ is chosen  to fit known ionization potential (IP) of the atom. In HF approximation IP is 
just the energy of outermost electron of the atom, IP$=\min(|\epsilon_c|)$. Note that parameter $a$ in 
(\ref{e:Vp}) has the same meaning and should be close in value to the atomic polarizabily $\alpha_0$ in
(\ref{e:pol}). However, in our calculations they are significantly different due to approximate treatment of
the correlations. Equation (\ref{e:pol}) is only valid on large distances where it has no effect on atomic
orbitals, while we use (\ref{e:Vp}) on all distances. On short distances real correlation potential is very
complicated, non-local, and cannot be reduces to either (\ref{e:pol}) or (\ref{e:Vp}).

\begin{table*}
\caption{\label{t:polc}
Ionization potentials (IP) and static scalar  polarizabilities $\alpha_0$ for closed-shell atoms, 
including superheavy elements E112, E118 and E120. Comparison with other calculations and experiment. 
All numbers are in atomic units.} 
\begin{ruledtabular}
\begin{tabular}{cll dc rdcdl}
&&\multicolumn{1}{c}{Ground}&
\multicolumn{5}{c}{Present work}&\multicolumn{2}{c}{Other} \\
\multicolumn{1}{c}{$Z$}&
\multicolumn{1}{c}{Atom}&
\multicolumn{1}{c}{State}&
\multicolumn{1}{c}{IP$(a=0)$}&
\multicolumn{1}{c}{$\alpha_{0}(a=0)$}&
\multicolumn{1}{c}{$a$}&
\multicolumn{1}{c}{IP}&
\multicolumn{1}{c}{$\alpha_0$}&
\multicolumn{1}{c}{IP}&
\multicolumn{1}{c}{$\alpha_0$} \\
\hline
%Z  Atom IP(exp)   IP      polc     a    polc      Other
\multicolumn{10}{c}{Noble gases}\\
 36 & Kr  & $4s^24p^6$ & 0.514 & 16.5 &  0 & 0.514  & 16.5 & 0.514\tablenotemark[1] & 16.47 - 17.21\tablenotemark[2] \\
 54 & Xe  & $5s^25p^6$ & 0.440 & 27.0 & 14 & 0.446  & 26.7 & 0.446\tablenotemark[1] & 26.97 - 28.22\tablenotemark[2]  \\
 86 & Rn  & $6s^26p^6$ & 0.384 & 35.0 & 27 & 0.395  & 34.2 & 0.395\tablenotemark[1] &  33.18\tablenotemark[3] \\

\multicolumn{10}{c}{$ns^2$-elements}\\                

 56 & Ba   & $5p^{6}6s^2$   &  0.163 & 324  & 142  & 0.192 & 251   & 0.192\tablenotemark[1]  & 272 - 275\tablenotemark[2] \\
 70 & Yb   & $4f^{14}6s^2$  &  0.197 & 178  & 125  & 0.230 & 142   & 0.230\tablenotemark[1]  & 139 - 141\tablenotemark[2] \\
 80 & Hg   & $5d^{10}6s^2$  &  0.328 & 44.9 & 135  & 0.384 & 39.1  & 0.384\tablenotemark[1]  & 31.32 - 33.91\tablenotemark[2] \\
 88 & Ra   & $6p^{6}7s^2$   &  0.166 & 297  & 138  & 0.194 & 232   & 0.194\tablenotemark[1]  & 248.56\tablenotemark[4] \\

\multicolumn{10}{c}{Superheavy elements}\\	                

102 & No    & $5f^{14}7s^2$ &  0.209  & 143 &  125  & 0.245 & 114 &  0.245\tablenotemark[5] &  110\tablenotemark[5] \\
112 & Cn    & $6d^{10}7s^2$ & 0.442  & 30.5 &  80  & 0.480 & 29.0 &  0.440\tablenotemark[6] &  27.40\tablenotemark[6] \\
112 & Cn    & $6d^{10}7s^2$ &           &      & 135  & 0.507 & 28.2 &  &   \\
118 & E118  & $7s^27p^6$    & 0.306  & 61.0 &  27  & 0.315 & 59.0 &  0.327\tablenotemark[7]& 46.3\tablenotemark[7]; 52.4\tablenotemark[8] \\
118 & E118  & $7s^27p^6$    &            &      &  54  & 0.325 & 57.2 &  &    \\
120 & E120  & $7s^27p^68s^2$& 0.192  & 184  &  140 & 0.234 & 147  & 0.215\tablenotemark[9] & 163\tablenotemark[9]  \\

\end{tabular}
\end{ruledtabular}
\tablenotetext[1]{Experimental IP from NIST database~\cite{NIST}.}
\tablenotetext[2]{Range of most accurate theoretical and experimental results from compilation of Ref.~\cite{mitroy2010}}
\tablenotetext[3]{Theory, Ref.~\cite{nakajima2001}.}
\tablenotetext[4]{Theory, Ref.~\cite{lim2004}.}
\tablenotetext[5]{Theory, Ref.~\cite{DSS14}.}
\tablenotetext[6]{Theory, Ref.~\cite{pershina2008p}.}
\tablenotetext[7]{Theory, Ref.~\cite{pershina2008a}.}
\tablenotetext[8]{Theory, Ref.~\cite{Nash2005}.}
\tablenotetext[9]{Theory, Ref.~\cite{borschevsky2013ab}.}
%\end{ruledtabular}
\end{table*}
%----------------------------------------------------------------------------------------------------------------------------------------------

Table~\ref{t:polc} presents results of the calculations for some closed-shell atoms including superheavy elements 
Cn, E118 and E120. As one can see the RPA approximation (\ref{e:HFDPsi}) works very well for Noble gases. 
Note that we present for comparison only range of most accurate earlier experimental and  theoretical results.
Detailed review of atomic polarizabilities can be found in Ref.~\cite{mitroy2010}, while the aim of the current
comparison is to check the validity of the method.
Fitting the IP with the polarization potential (\ref{e:Vp}) has little effect on the polarizability of Xe and moves
polarizability of Rn closer to the result of more sophisticated calculations of Ref.~\cite{nakajima2001}. 
The superheavy element E118 is the heaviest noble gas atom.  It is natural to expect some similarities
in the fitting procedure and accuracy of the results. Given that the value of fitting parameter $a$ 
(see formula (\ref{e:Vp})) increases two times from Xe ($a=14$) to Rn ($a=27$) we performed the 
calculations for E118 at two values of the parameter, one as in Rn ($a=27$) and another is two
times larger. Resulting IP of E118 at $a=54$ is in excellent agreement with the results of more
accurate calculations of Ref.~\cite{pershina2008a}. However, the polarizabilities are
significantly different (57.2 and 46.3, see Table~\ref{t:polc}). The reason for this difference is not
clear. Note, that the agreement with calculations of Ref.~\cite{Nash2005} is noticeably better.

The next group of atoms is represented by Ba, Ra and E120. Here correlations are large.
This is probably because the two $s$-electrons start a new shell which is well separated 
from the core electrons both in space and on energy scale. It is interesting that fitting of the IP for Ba and Ra
is achieved at very close values of the parameter $a$ ($a=142$ for Ba and $a=138$ for Ra).
Resulting polarizability is 7-8\% smaller than in accurate calculations (see Table~\ref{t:polc}).
Assuming similarity between Ba, Ra and E120 we perform calculations for E120 at $a=140$ a.u.
and resulting value of polarizability ($\alpha_0 = 147$ a.u.) increase by 8\%. The result,
$\alpha_0 = 159$ a.u., is in excellent agreement with accurate calculations of Ref.~\cite{borschevsky2013ab}.
Final value of $\alpha_0$ for E120 is presented in Table~\ref{t:final}.

The method works surprisingly well for ytterbium and nobelium. One value of the fitting parameter $a$ 
($a=135$ a.u.)  fits IP and polarizabilities of both atoms (see Table~\ref{t:polc}). This is important because
accurate calculations for Yb are difficult due to the role of excitations from the $4f$ subshell. This even
led to significant disagreement between calculations of the polarizabilities of Yb by different groups
(see Ref.~\cite{DD-Yb10} for details). It turns out that correct value of the polarizability can be obtained in 
a relatively simple way treating excitations from the $4f$ shell perturbatively~\cite{DD-Yb10}. 
This opens a way of calculating polarizabilities for atoms with open $4f$ or $5f$ subshell
(see next section).

In contrast to Yb and No, Hg and Cn atoms present the least accurate case. Fitting of the energy 
leads to polarizability which is about 15\% larger than more accurate values from other sources
(see Table~\ref{t:polc}). However, even in this case the polarizability of Cn agrees pretty well
the the more accurate calculations of Ref.~\cite{pershina2008p}.

\section{Open-shell atoms}

In this section we consider atoms which have two $s$ valence electrons and open $d$ or $f$ shells.
Corresponding configurations are $4f^n6s^2$, $5d^n6s^2$ and $6d^n7s^2$.  Calculations are done
in a way very similar to what is described in previous section. The only difference is the use of fractional
occupation numbers. The contribution of the open shells to HF potential and polarizability is multiplied 
by the fraction $x_d = n_d/10$ or $x_f = n_f/14$, where $n_d$ is the actual number of electrons on the
$d$-shell and $n_f$ is the actual number of electrons on the $f$-shell. For atoms with two $s$-electrons 
on the outermost shell polarizabilities dominated by contributions from these $s$-electrons which is
treated accurately. Contribution from the open $d$ of $f$-shell is smaller, but it is also treated pretty
well at least for the case of almost filled shell ($n_d \sim 10$ or $n_f \sim 14$).

\begin{table*}
\caption{\label{t:pol0}
Ionization potentials and static scalar  polarizabilities $\alpha_0$ of some atoms with open $4f$ or $5d$ 
subshells. Comparison with other calculations and experiment. 
All numbers are in atomic units.} 
\begin{ruledtabular}
\begin{tabular}{cll ddc rccc}
&&&\multicolumn{1}{c}{Expt.}&
\multicolumn{4}{c}{Present work}&\multicolumn{2}{c}{Other theory, $\alpha_0$} \\
\multicolumn{1}{c}{$Z$}&
\multicolumn{1}{c}{Atom}&
\multicolumn{1}{c}{Ground State}&
\multicolumn{1}{c}{IP}&
\multicolumn{1}{c}{IP$(a=0)$}&
\multicolumn{1}{c}{$\alpha_{0}(a=0)$}&
\multicolumn{1}{c}{$a$}&
\multicolumn{1}{c}{$\alpha_0$}&
\multicolumn{1}{c}{\cite{DKF14}}&
\multicolumn{1}{c}{\cite{Doolen}} \\
\hline
%Z  Atom IP(exp)   IP      polc     a    polc      Other
\multicolumn{9}{c}{$4f$-elements}\\                

 66 & Dy  & $4f^{10}6s^2$ & 0.218 &  0.188 & 209  &119  &168  & 162.7 & 165  \\
 67 & Ho  & $4f^{11}6s^2$ & 0.221 &  0.190 & 201  &121  &161  & 156.3 &  159 \\
 68 & Er   & $4f^{12}6s^2$ & 0.224 &  0.192 & 193  &123  &154  & 150.2 & 153 \\
 69 & Tm  & $4f^{13}6s^2$ & 0.227 &  0.194 & 186  &125  &147  & 144.3 & 147 \\

\multicolumn{9}{c}{$5d$-elements}\\	                

 73  &Ta   & $5d^36s^2$ & 0.277  & 0.254  & 81.4  & 75 &  73.7  &&            88.4 \\
 74 & W   & $5d^46s^2$ & 0.289  & 0.267  & 74.4 &  68  & 68.1  &&            74.9 \\
 75 & Re  & $5d^56s^2$ &  0.288  & 0.278  & 67.9  & 28  & 65.6  &&            65 \\
 76 & Os  & $5d^66s^2$ & 0.310  & 0.289  & 62.2  & 58  & 57.8  &&           57 \\
 77 & Ir    & $5d^76s^2$ & 0.330  & 0.299  & 57.0  & 83  & 51.7 &  54.0(6.7)\tablenotemark[1] & 51 \\
	                
\end{tabular}
\tablenotetext[3]{Experiment, Ref.~\cite{bardon1984polarizability}.}
\end{ruledtabular}
\end{table*}

Table \ref{t:pol0} shows calculated IP and polarizabilities of some lanthanoids and atoms with open $5d$
subshell. One can see that fitting of the IPs for lanthanoids with polarization potential (\ref{e:Vp}) leads to the
values of polarizabilities which are in very good agreement with earlier much more sophisticated and
presumedly more accurate results of Ref.~\cite{DKF14} and with old results of 
Doolen from the Muller's handbook~\cite{Doolen}. The value of the fitting parameter $a$ changes 
little from atom to atom going down
monotonically with reducing number of $f$-electrons. For the heaviest of the considered atoms, Tm, the
value of $a$ is the same as for its neighbour in the periodic table, Yb (see Table~\ref{t:polc} in previous section).
Good agreement between results of present work and those of Ref.~\cite{DKF14} adds to the validity of both 
methods. However, we beilive  that the results of \cite{DKF14} are more accurate, therefore, we don't consider 
any more atoms with open $4f$ or $5f$ subshell.

Next we consider atoms with open $5d$ shell. The purpose is to find the values of fitting parameter $a$ to use 
in the calculations for superheavy elements from Db to Cn. We do not consider Lr and Rf as well as their lighter 
analogs Lu and Hf because these atoms have relatively simple electron structure and accurate calculations for
them were performed in our earlier work~\cite{DSS14}. Among atoms with open $5d$-shell we
consider only those which have the $5d^n6s^2$ configuration in the ground state. The results are presented
in Table ~\ref{t:pol0}. The values of polarizabilities are in good agreement with earlier calculations~\cite{Doolen}
and with experiment for Ir~\cite{bardon1984polarizability}.

\begin{table*}
\caption{\label{t:pols}
Ionization potentials and static scalar  polarizabilities $\alpha_0$ of superheavy atoms with open 
$6d$ subshell.  Results of other calculations for IP are also presented. All numbers are in atomic units.} 
\begin{ruledtabular}
\begin{tabular}{ccl ddddd}
&&\multicolumn{1}{c}{Ground}&
\multicolumn{2}{c}{$a=0$}&
\multicolumn{2}{c}{$a=80$}&
\multicolumn{1}{c}{Other~\cite{türler2013}}\\
\multicolumn{1}{c}{$Z$}&
\multicolumn{1}{c}{Atom}&
\multicolumn{1}{c}{State}&
\multicolumn{1}{c}{IP}&
\multicolumn{1}{c}{$\alpha_{0}$}&
\multicolumn{1}{c}{IP}&
\multicolumn{1}{c}{$\alpha_{0}$}&
\multicolumn{1}{c}{IP}\\
\hline
%Z  Atom IP(exp)   IP      polc     a    polc      Other
%           IP    polc   a    IP    polc
105 & Db  & $6d^37s^2$    &  0.218 & 45.6  &  0.248 & 42.5 & 0.271 \\
106 & Sg  & $6d^47s^2$    &  0.252 & 43.9  &  0.283 & 40.7 & 0.258; 0.288 \\
107 & Bh  & $6d^57s^2$    & 0.284 & 41.3  &  0.317 & 38.4 &  0.251; 0.28\\
108 & Hs  & $6d^67s^2$    & 0.316 & 38.8  &  0.350 & 36.2 & 0.246; 0.28 \\
109 & Mt  & $6d^77s^2$    & 0.348 & 36.4  &  0.383 & 34.2 & 0.32 \\
110 & Ds  & $6d^87s^2$    &  0.379 & 34.2  &  0.415 & 32.3 & 0.35 \\
111 & Rg  & $6d^97s^2$    &  0.411 & 32.3  &  0.448 & 30.6 & 0.390 \\
%112 & Cn  & $6d^{10}7s^2$ & 0.442 & 30.5  &  0.480 & 29.0 \\

\end{tabular}
\end{ruledtabular}
\end{table*}

Table~\ref{t:pols} shows the results of the calculations of IP and polarizabilities of superheavy elements 
from Db to Rg. Note that in contrast to their lighter analogs all these atoms have the $6d^n7s^2$ ground 
state configuration. This is due to well known relativistic contraction of the $7s$ orbital (see, e.g. \cite{eliav2015}).
We use the largest value of the fitting parameter $a$ ($a=80$ a.u.) among those found for lighter atoms
(see Table~\ref{t:pol0}). Comparing results with $a=0$ and $a=80$ shows that contribution of correlations 
is relatively small and its approximate treatment is therefore justified. 
This difference also gives some idea about the accuracy of the calculations.

Last column of Table~\ref{t:pols} shows the results of other calculations of the IP of superheavy elements.
The numbers are taken from the review paper~\cite{türler2013}. When two numbers are presented, the
first number is the result of multi-configurational Dirac-Fock calculations and the second number is
corrected IP obtained with an extrapolation procedure (see Re.~\cite{türler2013} for details and for 
references to original works). There is excellent agreement between our calculations and those presented
in \cite{türler2013} for Sg. For other atoms the difference is larger and varies between 10 and 20\%.
We take into account this difference in estimating the uncertainty of present calculations.

\begin{table}
\caption{\label{t:final}
Final values of ionization potentials and polarizabilities of superheavy elements, 
including uncertainies.  All numbers are in atomic units.} 
\begin{ruledtabular}
\begin{tabular}{ccl cc}
\multicolumn{1}{c}{$Z$}&
\multicolumn{1}{c}{Atom}&
&%\multicolumn{1}{c}{State}&
\multicolumn{1}{c}{IP}&
\multicolumn{1}{c}{$\alpha_{0}$}\\
\hline
%Z  Atom IP(exp)   IP      polc     a    polc      Other
%           IP    polc   a    IP    polc
105 & Db  & $6d^37s^2$    &  0.25(3) & 42(4)   \\
106 & Sg  & $6d^47s^2$    &  0.28(3) & 40(4)   \\
107 & Bh  & $6d^57s^2$     & 0.32(4) & 38(4)   \\
108 & Hs  & $6d^67s^2$     & 0.35(4) & 36(4)   \\
109 & Mt  & $6d^77s^2$     & 0.38(4) & 34(3)   \\
110 & Ds  & $6d^87s^2$     &  0.41(4) & 32(3)   \\
111 & Rg  & $6d^97s^2$     &  0.45(4) & 30(3)  \\
112 & Cn  & $6d^{10}7s^2$ & 0.48(4) & 28(4)   \\
118 & E118  & $7s^{2}7p^6$ & 0.33(3) & 57(3)   \\
120 & E120  & $7s^27p^28s^2$ & 0.23(4) & 159(10)   \\

\end{tabular}
\end{ruledtabular}
\end{table}

Final results for all superheavy elements considered in present paper are summarised in Table~\ref{t:final}.
Here the results include uncertainty which is estimated from the analysis of the data in previous tables.
Note that accurate calculations are available for Cn~\cite{pershina2008p}, E118~\cite{pershina2008a} 
and E120~\cite{borschevsky2013ab} due to their simple electron structure.
We include the results for this atoms to illustrate that reasonablely accurate results can be obtained 
with a very simple method.

\acknowledgments

The author is thankful to Valeria Pershina, Michael Block and Peter Schwerdtfeger for stimulating discussions.
The work was supported in part by the Australian Research Council.

\bibliographystyle{apsrev}

\end{document}